# Distinguishing antiferromagnetic spin sublattices via the spin Seebeck effect

Yongming Luo[1,2], Changjiang Liu[2], Hilal Saglam[2], Yi Li[2], Wei Zhang[4], Steven S.-L. Zhang[2,5], John E. Pearson[2], Brandon Fisher[2], Anand Bhattacharya[2*] and Axel Hoffmann[2,3*]

[1] School of Electronics and Information, Hangzhou Dianzi University, Hangzhou, Zhejiang, 310018, China.

[2] Materials Science Division, Argonne National Laboratory, Argonne, Illinois 60439, USA

[3] Department of Materials Science and Engineering, University of Illinois at Urbana-Champaign, Urbana, Illinois 61801, USA

[4] Department of Physics, Oakland University, Rochester, MI 48309, USA

[5] Department of Physics, Case Western Reserve University, Cleveland, Ohio 44106, USA

*Yongming Luo and Changjiang Liu contribute equally to this work*

*Corresponding to Axel Hoffmann (axelh@illinois.edu) and Anand Bhattacharya (anand@anl.gov)

Antiferromagnets are beneficial for future spintronic applications due to their zero magnetic moment and ultrafast dynamics. But gaining direct access to their antiferromagnetic order and identifying the properties of individual magnetic sublattices, especially in thin films and small-scale devices, remains a formidable challenge. So far, the existing read-out techniques such as anisotropic magnetoresistance, tunneling anisotropic magnetoresistance, and spin-Hall magnetoresistance, are even functions of sublattice magnetization and thus allow us to detect different orientations of the Néel order for antiferromagnets with multiple easy axes. In contrast direct electrical detection of oppositely oriented spin states along the same easy axes (e.g., in uniaxial antiferromagnets) requires sensitivity to the direction of individual sublattices and thus is more difficult. In this study, using spin Seebeck effect, we report the electrical detection of the two sublattices in a uniaxial antiferromagnet $Cr_2O_3$. We find the rotational symmetry and hysteresis behavior of the spin Seebeck signals measured at the top and bottom surface reflect the dierction of the surface sublattice moments, but not the Néel order or the net moment in the bulk. Our results demonstrate the important role of interface spin sublattices in generating the spin Seebeck voltages, which provide a way to access each sublattice independently, enables us to track the full rotation of the magnetic sublattice, and distinguish different and antiparallel antiferromagnetic states in uniaxial antiferromagnets.

Antiferromagnetic materials are promising for future spintronic applications for a wide variety of reasons [1,2,3,4]. For example, they generate zero stray fields, which automatically eliminate unintentional magnetic crosstalk between neighboring devices, and additionally makes devices robust against perturbations from magnetic fields. This is directly beneficial for high stability, high density data storage [5]. They also exhibit intrinsic high frequency dynamics [6] and may be used as controllable THz resonators [7,8]. Recent experiments show that the antiferromagnetic Néel order can be manipulated by current induced torques, such as spin-galvanic effects [9,10], interfacial spin-orbit torques [11,12], or by electrical voltage using magnetoelectric effects [13,14]. Antiferromagnets also can have a large spin Hall angle, which can be used for effective conversion of spin current and charge current [15,16], and realize field free switching of a ferromagnet when combined with exchange bias [17,18]. In addition, antiferromagnets are also promising for magnon based devices [19,20,21], where information is processed and transmitted by magnons. Recent experiments show that antiferromagnetic insulators can also generate magnons by thermal spin-wave excitation [22,23], where the magnons in antiferromagnetic insulator can propagate over distances of more than tens of µm [24,25]. Theory also predicts that antiferromagnets may support spin super-fluidity [26]. These results point to a more active role for antiferromagnets in high stability, ultrafast, and low-power antiferromagnetic spintronic devices.

However, despite the rapidly growing amount of research focused on antiferromagnetic spintronics, the field is still in its infancy for real practical application. The development of methods for reliable and high signal-to-noise read-out of antiferromagnetic states [1, 2] remains an outstanding challenge within this field. Recent studies show that antiferromagnetic order can be detected by magneto-transport measurements: such as anisotropic magnetoresistance [5], tunneling anisotropic magnetoresistance [27], and spin-Hall magnetoresistance [28,29,30,31,32]. However, these effects are even functions of the magnetic order, and thus are only sensitive to the reorientation of the Néel order along different easy axes, requiring multiple contacts for generating currents in different orientations, which limits device miniaturization. Detection of opposite antiparallel spin arrangements in uniaxial antiferromagnets may overcome this obstacle.

Spin Seebeck effect (SSE), where a temperature gradient results in a spin current $\vec{J_s}$ carrying spin angular momentum $\vec{\sigma}$, flowing along $\vec{\nabla T}$. When a paramagnetic metal with large spin orbit coupling (such as Pt, W) is present at the surface, the spin current will inject into it and subsequently converted into a measurable voltage through the inverse spin Hall effect. The associated voltage $V_{SSE}$ is given by:

$$V_{SSE} \propto L_v \theta_{SH} (\vec{J}_s \times \vec{\sigma}) \tag{1}$$

Here, $\theta_{SH}$ is the spin Hall angle of the heavy metal layer, is the spin current (the direction of vector is its flow direction) and $L_v$ is the separation between the electrodes for the voltage measurement [33,34,35]. According to existing theories for the SSE, SSE may be surface sensitive and may reflect the direction of the surface magnetic moments. This is because the polarization direction of the spin current is determined by the dynamical moment of the magnons that carry the heat current, which may be modulated by the surface moments that are in contract with the the paramagnetic material. The pumping of spin current into paramagnetic layer is due to the interfacial exchange coupling between the conduction electrons in paramagnetic metal and magnetic moments in magnetic materials, and only the magnons that are coupled to the interface moment can be converted into spin currents [36]. The coupling length scale, which is determined by the overlap of Fermi wavelength, is at atomic scale (~1 nm for Pt/YIG system [37]). SSE provide a new way to detect the order in antiferromagnetic materials and recent studies have observed the SSE in antiferromagnetic materials via a sudden change of the SSE signals at a spin flop transition [22, 23]. However, the exact role of the surface sublattices in generating SSE signals has remained unclear.

In this manuscript, we measured the SSE signals at the top and bottom surface of a uniaxial antiferromagnetic material $Cr_2O_3$. Our results demonstrate the role of the surface moment in generating spin Seebeck voltages, which not only provide an important insight for validating theoretical models, but also provide a way to access the individual spin sublattices. We can track the full rotation of the two magnetic sublattices and the different antiparallel spin states in uniaxial antiferromagnets can be distinguished.

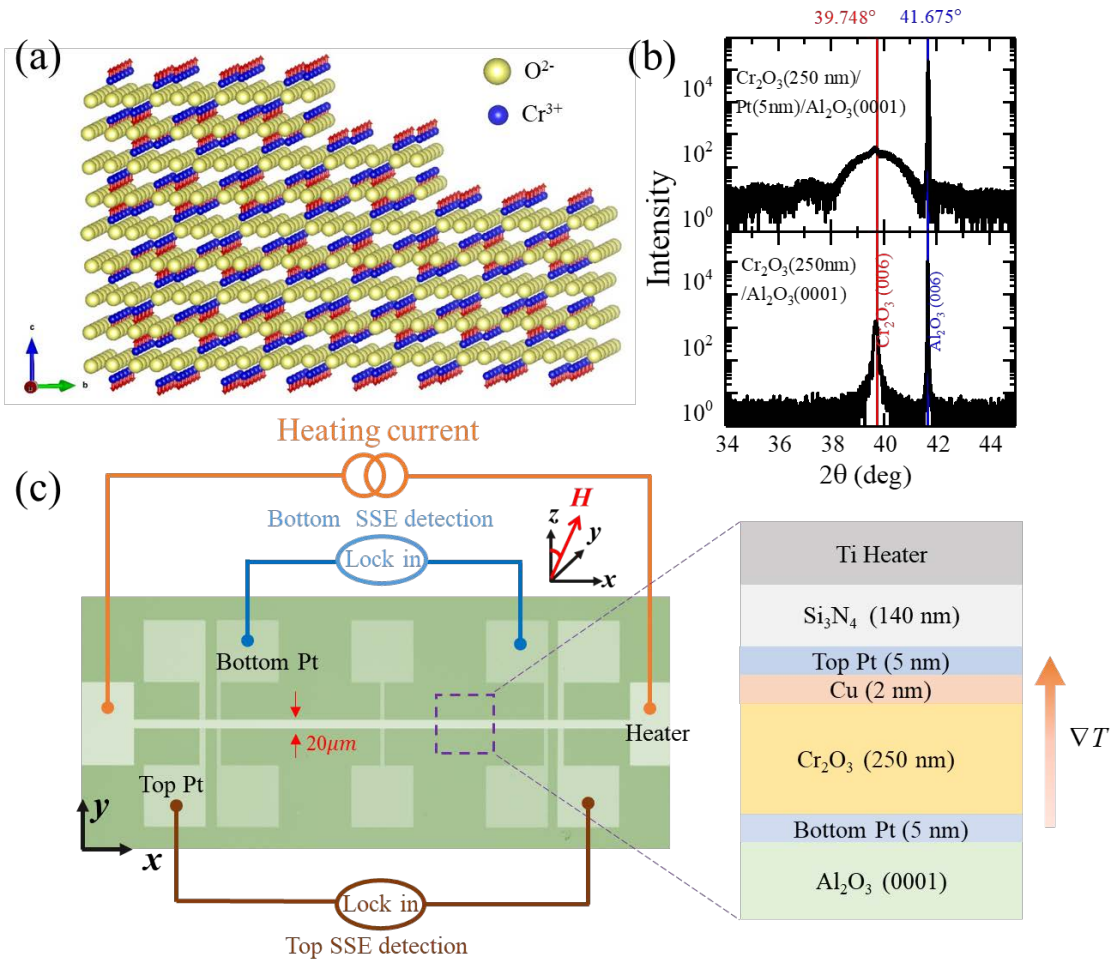

Fig.1: (a) Illustration of the spin structure of a $Cr_2O_3$ single crystal with a stepped (0001) surface. The red arrows point along the $c$ axis denote the spin direction of $Cr^{3+}$ ions. (b) $\theta$-$2\theta$ X-ray diffraction pattern of a 250-nm $Cr_2O_3$ film (upper panel) and $Cr_2O_3$ (250 nm)/Pt(5 nm) bilayer (lower panel), respectively. The films are deposited on $Al_2O_3$ (0001) substrates. (c) Experimental set up of the SSE measurements. The insert on the right shows the cross section of the sample structure.

**Experimental set up**

Cr$_2$O$_3$ is a unique uniaxial antiferromagnetic material. Unlike conventional antiferromagnets, where the moment of an uncompensated surface usually averages out due to surface roughness, the Cr$_2$O$_3$ (0001) surface exhibit long-range magnetic order, even in the presence of surface roughness [38]. Figure 1(a) illustrates a configuration of the Cr$_2$O$_3$ (0001) surface. For a particular antiferromagnetic domain, the surface Cr$^{3+}$ ion moments are parallel aligned, even with surface roughness. Besides, the Cr$^{3+}$ spins at the top and bottom surface are always opposite. The antiparallel moments at top and bottom surface can represent two different spin sublattices of the Cr$_2$O$_3$. Such unique spin structure is due to the requirements of charge-neutrality and the nature of interlayer antiferromagnetic coupling in crystalline Cr$_2$O$_3$ [39]. The existence of different sublattices at top and bottom surfaces make it possible to access individual sublattice, and distinguish different antiparallel states in uniaxial antiferromagnetic material, by using surface sensitive techniques.

To detect the spin current at the top and bottom surfaces of Cr$_2$O$_3$, we grow Pt layers on the top and bottom surface of the Cr$_2$O$_3$ film. Here we choose Pt is because it is a well-studied spin current detector with large spin orbit coupling, and Cr$_2$O$_3$ can grow epitaxially on Pt [39]. We first characterize the film crystal structure. Figure 1(b) shows the X-ray diffraction (XRD) of a 250 nm Cr$_2$O$_3$ single layer grown on Al$_2$O$_3$ (0001) substrate with and without Pt bottom layer. The red and blue line are the standard (0006) peak positions for Cr$_2$O$_3$ and Al$_2$O$_3$, respectively. For the Cr$_2$O$_3$ film grow on Al$_2$O$_3$ (0001), a clear diffraction peak appears at the (0006) peak position. For Cr$_2$O$_3$ grow on Pt, we can still observe the tip at the Cr$_2$O$_3$ (0006) Bragg reflection on top of the broad intensity oscillations arising from the thin Pt layers. The comparison of XRD results with Cr$_2$O$_3$ (0006) peak between film grown on Pt and grown directly grown on the substrate, demonstrates the *c*-axis oriented crystal structure of Cr$_2$O$_3$ on Pt.

FIG. 1 (c) summarize sample structure and experiment set up. We grow a Pt (5 nm)/Cu(2 nm)/$Cr_2O_3$(250 nm) /Pt(5 nm) layer stack. Here the insert of 2-nm Cu between the top Pt/$Cr_2O_3$ interface is to reduce the induced moment in Pt due to the proximity effect [40]. The films are patterned into 600 μm × 20 μm Hall-bar structures for transport measurement, as shown in the left panel of Fig. 1(c). In addition, in order to provide a temperature gradient during the SSE measurements, we fabricate an on-chip heating structure. A 140-nm electrical insulating $Si_3N_4$ layer, and a resistive Ti layer was deposited on top of the layer stack to serve as a resistive heater for the device. The right panel Fig. 1(c) shows cross section of the sample structure. The detail sample fabrication is shown in methods.

During the measurement, we send a sinusoidal current through the top Ti heater layer ($P_{heating}$ ~ 5 $mW_{rms}$), which generate a temperature gradient $\vec{\nabla}T$ normal to the film plane. The resulting $V_{SSE}$ at the top and bottom Pt layers are measured using lock-in techniques at the second harmonic, as shown in Fig.1 (c). As $\vec{\sigma}$ is coupled to the magneic moment in $Cr_2O_3$, to be accurate, the *y* component of the moment in our longitudinal SSE measurement geometry, we apply magnetic fields with different amplitudes in the *y-z* plane, as shown in the inset of Fig 1. (c).

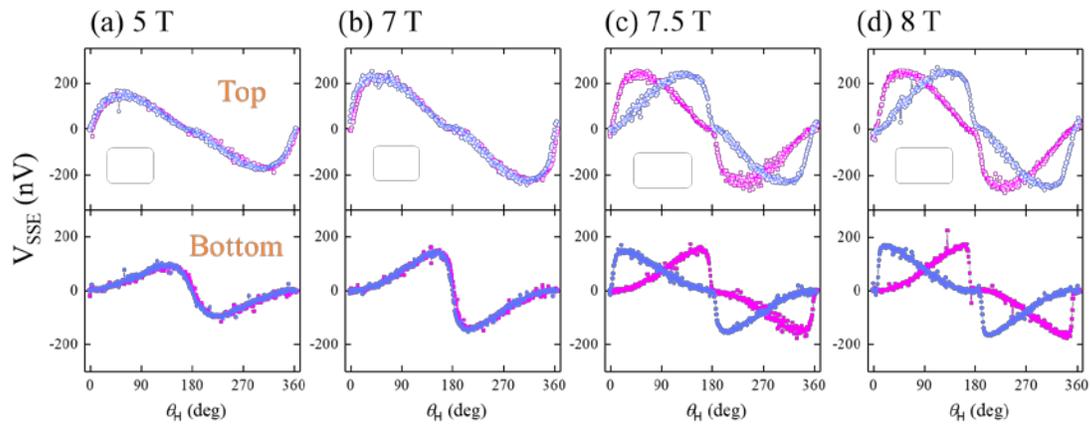

Fig. 2 Angular dependence of the SSE voltages at the top and bottom surface of $Cr_2O_3$ at different fields. In (a) (b) (c) and (d), The upper figure with hollow dot lines represent the signals measured

at the top Pt layer, while the lower figure with solid dot lines represent the signals measured at the bottom Pt layer. The magenta and light blue colors represent the signals when the field rotate 0° → 360° and rotate back 360° → 0°, respectively. $\theta_H$ is defined as in Fig. 1(c). The temperature during these measurements was 100 K.

**Experimental results**

Figure 2 shows the angular dependence of the spin Seebeck signals at different applied magnetic fields at 100 K. We observe a transition at 7.5 T. For fields below 7.5 T, the SSE voltages at the top and bottom surfaces do not show hysteresis when the field is rotated from 0° to 360° and rotated back from 360° to 0°, while a clear hysteresis appears above that. The rotational symmetry also changes across the transition field. At fields below 7.5 T, the signals at the top and bottom are antisymmetric about $\theta_H = 180°$ °. While after the transition, the signals before and after $\theta_H = 180°$ exhibit a 180° phase shift and sign reversal. This transition indicates different magnetic behavior of $Cr_2O_3$ across the transition field. We will demonstrate that such transition of symmetry and hysteresis is due to the different magnetic behavior of $Cr_2O_3$ at fields below and above the spin-flop transition.

Besides the transition, another important observation is that the SSE signals at the top and bottom surfaces can be clearly distinguished. Both before and after the transition, the rotational SSE signals from the two surfaces exhibit different shapes, with maxima in their amplitudes realized in different field directions. This observation indicates the SSE signals at two surfaces have different origins.

The magnitude of the SSE signal we measured is directly proportional to the heating power (see supplemental material), suggesting the thermoelectric nature of the signal similar to those reported in FMI materials such as YIG [41]. Beside at 100 K, we also conduct the measurement at different temperatures, as shown in the supplemental material, we can observe similar transition behaviors, although with a variation of the critical transition field.

**Micromagnetic simulations of the field dependence of the spin sublattices**

To further understand the origin of SSE signals, we simulate the angular dependence of the magnetic state of $Cr_2O_3$ at different fields. Considering $Cr_2O_3$ is a layered antiferromagnet along the *c*-axis [(0001) plane], we model our system with 30 magnetic layers, which are numbered 1 to 30 from the bottom to the top, with spin ferromagnetically coupled in the same layer, while antiferromagnetically coupled between adjacent layers, as shown in Fig. 3(b). We choose standard magnetic parameters for the $Cr_2O_3$ material, with additional details discussed in the Methods section. The simulated spin flop field is 7.1 T, which is close to the spin flop field in our device (7.5 T). We simulate the angular dependence of the sublattice moments at fields below (6 T) and above the spin-flop transition (8 T), We plot the angular dependence of the in-plane magnetization component ($m_y$) at the top surface ($m_y^{i=30}$), bottom surface ($m_y^{i=1}$), and the net magnetization in the bulk $\sum_{i=1}^{30} m_y^i$, when magnetic fields are applied in the *y-z* plane. The magnetization configurations at selected magnetic field orientations are also plotted. The simulation results are shown in Fig. 3.

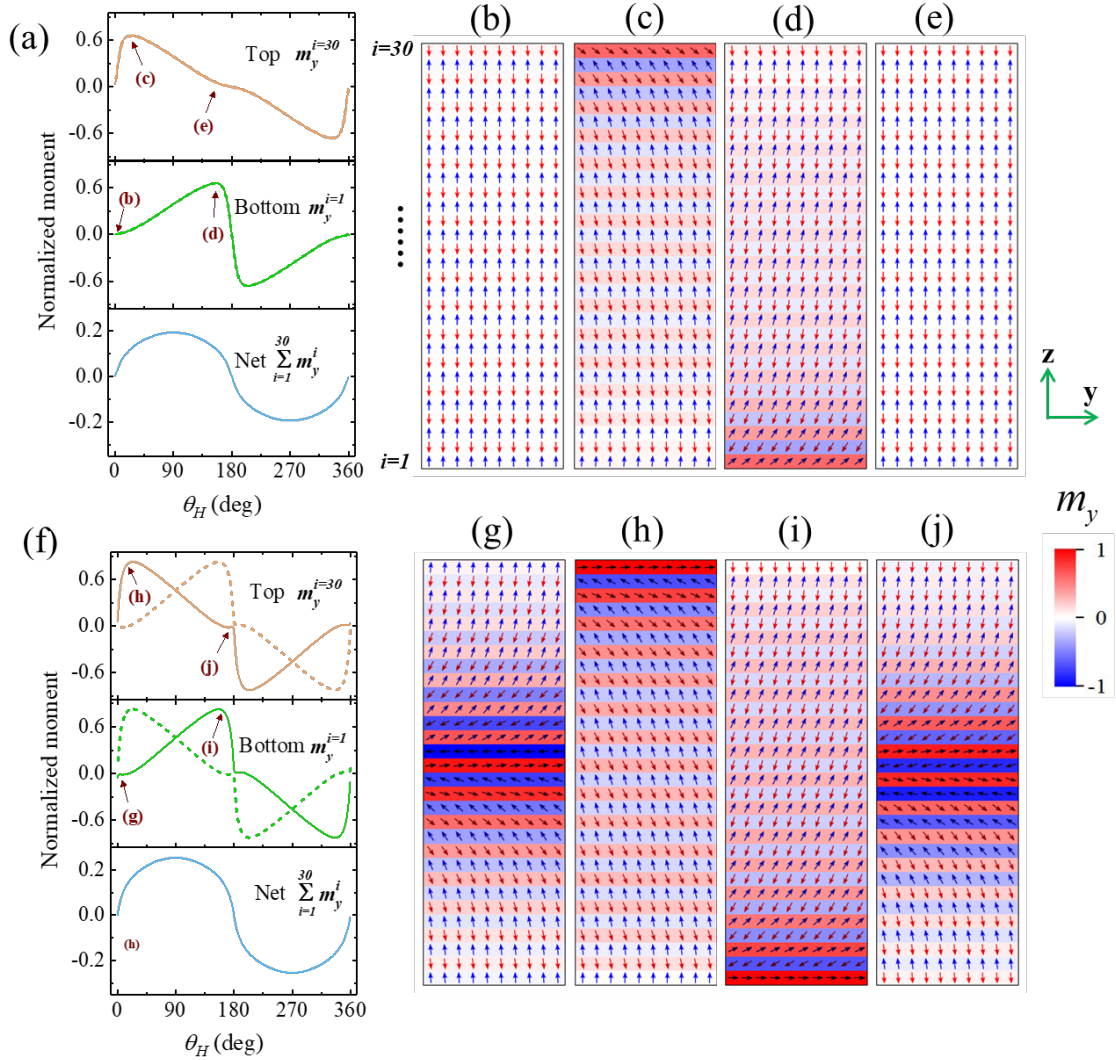

Fig. 3. Simulation of the angular magnetic field dependence of the sublattice moments at fields below (6 T) and above the spin flop field (8 T), as shown in (a) and (f), respectively. In (a) and (f), from top to down, are plots of the angular dependence of the normalized in-plane magnetization component at the bottom surface ($m_y^{i=1}$), top surface ($m_y^{i=30}$), and the net moment in the bulk ($\sum_{i=1}^{30} m_y^i$). The sold and dash lines represent the field rotate 0°→ 360° and rotate back 360°→0°, respectively. (b) – (e) and (g) – (j), show plots of the magnetization configuration at different selected field directions, indicated in (a) and (f). The small arrows indicate the sublattices in each layer. The colors denote the in-plane component as shown in the color bar. The number at the left side of the picture denote the layer number $i$.

Figure 3(a)-(e) show simulation results at field below spin flop transition (6 T). As shown in the top and middle plot in Figure 3(a), The angular dependences of top ($m_y^{i=1}$) and bottom ($m_y^{i=30}$) are non-hysteretic as the field is rotated clockwise (360°→ 0°) or counterclockwise (0°→360°), the line shapes are anti-symmetric about . Fig. 3(b) to (e) show the sublattice configurations at selected field directions, as denoted by the arrows in Fig. 3 (a). At initial state $\theta_H = 0°$, the two sublattices arrange antiparallel along easy axis. However, when the field rotate in the *y-z* plane and away from easy axis, the sublattices start to tilt from the easy axis and result in an in-plane component $m_y$, as shown in Figs. 3(b), (c), and (d), and the non-zero $m_y^{i=1}$ and $m_y^{i=30}$ in Fig. 3(a). This is due to the different magnetic susceptibility in antiferromagnets: the magnetic susceptibility perpendicular ($\chi_\perp$) to the Néel order is much larger than that parallel with Néel order ($\chi_\parallel$), i. e, $\chi_\perp \gg \chi_\parallel$ [42], so it is easier to induce magnetic moment when the field is not parallel with the Néel order. It should also be noticed that when the field is away from easy axis, the equilibrium position for two sublattices are different, the sublattice that is antiparallel with the *z* component of the field ($H_z$) will have larger in-plane component ($m_y$) than the one that is parallel to the field. Such difference is due to the opposite direction of the exchange field, anisotropy field that act on two antiparallel sublattices. This also result in different angular dependence line shapes of $m_y^{i=1}$ and $m_y^{i=30}$. The sublattices only tilt from the easy axis, but do not flip at fields below spin flop, as shown in Fig. 2(b) and (e), the magnetic configuration at 0° and 180° re the same. Such non-flip behavior is reversible and results in no hysteresis when the magnetic field rotates clockwise (0°→360°) and back counterclockwise (360°→0°).

The situation at fields above the spin flop transition is different. Figures 3(f)-(j) show simulation results at a field above the spin flop transition (8 T). As shown in Fig. 3(f), the angular dependences of $m_y^{i=1}$ and $m_y^{i=30}$ exhibit a 180° phase shift plus a sign reversal around $\theta_H = 180°$, and there is a clear hysteresis as the field rotates clockwise and rotate back counterclockwise. Figs. 3(g-j) show the detail sublattice behavior. When a larger magnetic field is applied along the easy axis, one observes a surface spin flop transition [43]. As shown in Fig. 3(g), the sublattices in the center of the system are in a flopped state, with large individual in-plane magnetization components, while for the sublattices at the top and bottom surfaces, they are parallel with the field. The gradual change of the sublattice direction from the center to the surface form an antiferromagnetic domain wall structure that located at the center. The domains at the different sides of the wall have different Néel orders, which are symmetric with the $z$ axis. When the field rotate in the $y$-$z$ plane, the collective behavior of the sublattices are manifest as the motion of the domain wall. Due to the difference in $\chi_\perp$ and $\chi_\parallel$ ($\chi_\perp \gg \chi_\parallel$), the domain with Néel order that has larger angle with respect to the field develops a moment in the field direction more easily and this determines the direction of the wall. As shown in fig. 3 (g) and (h), when the field rotate from 0° to 180°, the domain wall moves out from the top surface, then nucleate and return to the center from the bottom surface. During this process, both of the two sublattices flip their direction. As shown in Fig. 3 (j), all of the sublattices are flipped at $\theta_H = 180°$. The flip of the sublattices need to overcome the anisotropy and give rise to the hysteresis behavior when the field rotate clockwise (0°→360°) and rotate back (360°→0°).

A direct comparison of the simulation results in Fig. 3 with the SSE voltages in Fig. 2, we can conclude that the hysteresis and rotational symmetries of the SSE signals measured correspond to the top ($m_y^{i=30}$) and bottom ($m_y^{i=1}$) surface sublattice, the transition we observed at 7.5 T is due to different magnetic behavior at fields below and above the spin flop transition. The SSE signals do not correspond to net magnetic moment in the bulk ($\sum_{i=1}^{30} m_y^i$), which is symmetric at $\theta_H = 90°$ and $270°$, and exhibit no symmetry and hysteresis change at fields both below and above the spin flop transition. In the supplemental material, we also plot the angular dependence of the net moments of the two individual sublattices ($\sum_{i=1}^{15} m_y^{2i-1}$ and $\sum_{i=1}^{15} m_y^{2i}$), which also show different symmetry and hysteresis from the SSE signal.

Given that the SSE symmetry and hysteresis behavior are in accordance with the surface sublattice moment, Our results clearly show that the surface moments dominate the symmetry of the SSE signal, demonstrating that SSE is surface sensitive at the atomic scale. Our results emphasis the role of interfacial exchange coupling in generating the SSE signals. In antiferromagnetic materials with uncompensated surfaces, the sublattice at the uncompensated surface would have a stronger exchange coupling with the carriers in the detector layer, and so would dominate direction of SSE, since it sets the quantization axis for spin-flip scattering processes. Whenever the uncompensated surface spins lie in the plane and are perpendicular to direction of the SSE voltage electrodes, we detect a maximum in the SSE voltage. Given that our simulations suggest a significant rotation of the surface spins with respect to the bulk structure, the question then arises how angular momentum associated with the bulk magnon modes gets modified near the interface, and what role the excitations of the interface spins themselves play. Therefore, further theoretical and experimental work will be necessary to understand the heat current driven magnons and resulting spin currents in the interfacial region.

Our results provide a method to access the sublattices in antiferromagnets individually by using transport measurement. We can track the full rotation of the magnetic sublattice in real time, and can help to distinguish and read out different antiparallel spin states in uniaxial antiferromagnets. Fig. 4 show the angular dependence of the SSE signals for two different antiparallel spin states, at the field below spin flop transition (6 T). With different initial state, for the top and bottom surface, the rotational SSE signals show line shapes. The distinguishable line shapes help to distinguish the individual sublattice direction at the surfaces, and further the different antiparallel spin states in uniaxial antiferromagnets.

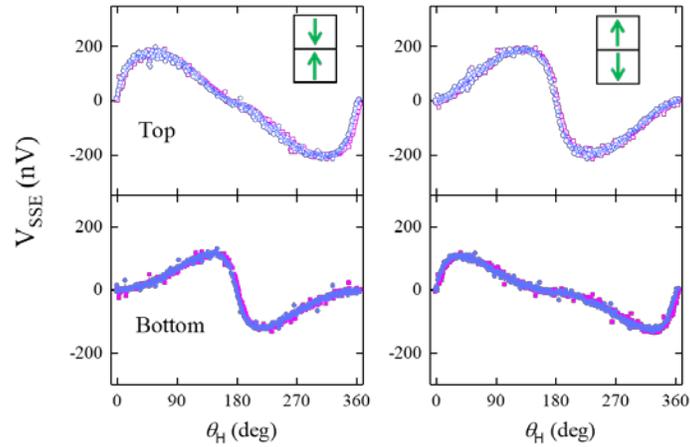

Fig.4 Distinguishing opposite sublattice configuration in uniaxial $Cr_2O_3$. (a) and (b) show the angular dependence of the SSE signals at the top and bottom surface with opposite initial sublattice directions, indicated by the insets. The temperature during the measurements was 100 K. The field amplitude is 6 T.

**Conclusion**

In conclusion, we measured the SSE at the top and bottom surface of $Cr_2O_3$ film. Our results demonstrate the important role of interface spin sublattices for generating the SSE voltages, which will be an important insight for validating theoretical models. Furthermore, we demonstrate that the SSE provides a way to directly detect the orientation of the spin sublattice at the uncompensated surface. This results emphas the surface sensitive property of SSE in reading out the antiferromagnetic state, which is different from the spin Hall magnetoresistance, anisotropy magnetoresistance or tunneling magnetoresistance, which are sensitive to the bulk magnetic order in antiferromagnetic materials.

**Method:**

**Sample fabrication:** We first deposit 5-nm Pt on an $Al_2O_3$ (0001) substrate by sputtering, and fabricated it into 600 µm × 20 µm bar structures by photolithography and ion milling. Then a 250-nm thick $Cr_2O_3$ layer was deposited on top by reactive sputtering at 600 °C. Subsequently, a 2-nm Cu and 5 nm Pt layer was deposited in-situ at room temperature. The top Pt layer was patterned into bar structures by photolithography and ion milling. Lastly, the on-chip heater was fabricated by using a lift-off process. The 100-nm $Si_3N_4$ layer was grown by PECVD, and the 30-nm Ti was deposited by sputtering.

**Simulation:** The simulation was conducted by using OOMMF. We simulate a system with 30 ferromagnetic layers, with spins that are ferromagnetically coupled (exchange stiffness $A_1 > 0$) in the same Cr layer, while antiferromagnetically coupled ($A_2 < 0$) between different layers, and with the same coupling strength $|A_1|=|A_2|$. The exchange energy was calculated with Neumann boundary conditions. We choose the standard parameters for $Cr_2O_3$: exchange stiffness $|A_1|=|A_2|=4\times10^{-12}\ J/m$, uniaxial anisotropy $K_u=2\times10^4\ J/m^3$, with the easy axis along the $z$ direction, and the magnetization of the sublattices $M_A=M_B=22579\ A/m$. The cell size in the simulation is $5\times5\times5\ nm$, and we simulate $10\times10$ cells in each layer. Our simulation shows that the critical field for spin-flop is 7.1 T, close to the spin-flop field in real bulk materials (6.5 T) [22]. In the simulation, the cell size is much large than the lattice constant in real $Cr_2O_3$ material, and thus it will influence the spin-flop field, but the angular magnetic field dependence of the sublattice moments below and above the spin-flop field is robust and independent of the different cell sizes.

**Acknowledgments**

This work was supported by the U. S. Department of Energy, Office of Science, Basic Energy Sciences, Materials Science and Engineering Division. We acknowledge José Holanda, Zhizhi Zhang and Valentine Novosad, for helpful discussion and help.

**Author Contributions**

Changjiang Liu designed the spin Seebeck sample structure. The Pt/$Cr_2O_3$/Pt trilayers were designed and grown by Yongming Luo, who also fabricated the spin Seebeck devices following Changjiang Liu's design using optical lithography and ion-milling. Changjiang

Liu made the experimental measurements that established the sensitivity of the spin Seebeck effect to the sublattice magnetizations. The data presented in this paper were obtained by Yongming Luo on structures that he fabricated. The micromagnetic simulations were performed by Yongming Luo. All authors were involved in the data analysis, discussion, and manuscript preparation. The project was supervised by Anand Bhattacharya and Axel Hoffmann.